\documentclass[prd,onecolum,preprintnumbers,showpacs,amsmath,amssymb,eqsecnum,nofootinbib]{revtex4}
\usepackage{mathpazo}
\DeclareSymbolFont{cmletters}{OML}{cmm}{m}{it} 
\DeclareMathSymbol{\xi}{\mathord}{cmletters}{"18}
\usepackage{graphicx}
\usepackage[dvipdfm]{hyperref}
\usepackage{dcolumn}
\usepackage{bm}
\preprint{DPNU-06-03}
\begin{document}
\title{Stability analysis of self-similar behaviors\\
 in perfect fluid gravitational collapse}

\author{Eiji Mitsuda}
\email{emitsuda@gravity.phys.nagoya-u.ac.jp}
\author{Akira Tomimatsu}
\email{atomi@gravity.phys.nagoya-u.ac.jp}
\affiliation{Department of Physics, Graduate School of Science, Nagoya
University, Chikusa, Nagoya 464-8602, Japan}
\begin{abstract} 
 Stability of self-similar solutions for gravitational collapse is an
 important problem to be investigated from the perspectives of their nature
 as an attractor, critical phenomena and instability of a naked singularity.
In this paper we study spherically symmetric
 non-self-similar perturbations of matter and metrics in spherically symmetric
 self-similar backgrounds.
The collapsing matter is assumed to be a perfect fluid with the equation
 of state $P=\alpha\rho$.
We construct a single wave equation governing the perturbations, 
which makes their time evolution in arbitrary self-similar backgrounds
 analytically tractable.
Further we propose an analytical application of this master wave equation to the
 stability problem by means of the normal mode analysis for the perturbations having the time dependence given by
 $\exp{(i\omega\log|t|)}$, and present some sufficient conditions for the absence of non-oscillatory unstable
 normal modes with purely imaginary $\omega$.

\end{abstract}
\maketitle
\section{Introduction\label{sec:intro}}
Gravitational collapse is a long-standing problem to be investigated in
general relativity.
Even if the analysis is restricted to a spherically symmetric system,
the Einstein equations still remain too complicated to understand fully
the generic features of the process of gravitational collapse.
One may try to overcome such a difficulty by considering self-similar
spherically symmetric spacetimes, for which the Einstein equations are
reduced to a set of ordinary differential equations with respect to the
one variable $z \equiv r/t$, where $r$ and $t$ may be comoving radial
and time coordinates.
Such self-similar solutions are useful for seeing interesting phenomena
in relation to the violation of the cosmic censorship due to naked
singularity formation and the critical behavior at the threshold of
black hole formation.
Further, as was advertised by Carr (see recent review \cite{CarrBJ:GRG37:2005}) as the so-called self-similarity
hypothesis, a self-similar behavior may be an attractor behavior, that
is, it may get dominant near a central dense region as gravitational collapse
starting from generic initial conditions proceeds to final stages.
Though this hypothesis strongly motivates us to study extensively
self-similar models, its validity should be confirmed through detailed
analysis of stability of self-similar behaviors for non-self-similar perturbations.

One approach to the stability problem may be to study time-evolution of
massless test (scalar or electromagnetic) fields in self-similar
background spacetime \cite{NolanBC:PRD66:2002, MitsudaE:PRD71:2005}.
In particular, the possible generation of burst-like emission of
electromagnetic radiation due to the infinite blue-shift effect at the
future Cauchy horizon associated with a central naked singularity was
discussed in \cite{MitsudaE:PRD71:2005} by means of the Green's function
technique.
Though the test-field approach is interesting for examining the
instability of the Cauchy horizon associated with a naked
singularity, our concern in this paper is rather to treat directly
non-self-similar perturbations of collapsing matter, which are accompanied
with metric perturbations.
To make our analysis tractable, we consider only spherically symmetric
perturbations. Further the collapsing matter is assumed to be a perfect
fluid with pressure $P$ given by the equation of state $P=\alpha\rho$,
where $\alpha$ is a constant in the range $0<\alpha\leq 1$.
This is because there exist various interesting classes of self-similar
spherically symmetric perfect fluid solutions (see
\cite{BJCarr:PRD62:2000} and \cite{CarrBJ:PRD67:2003} for example),  and their
stability has great implications to astrophysical problems.

In perfect fluid dynamics a sonic point located at $z=z_{s}<0$, whose
value is dependent on the equation of state parameter $\alpha$, plays the
role of a characteristic surface, instead of the null Cauchy surface
associated with the central singularity for
massless test fields.
Physically allowable self-similar solutions are constrained by
regularity (namely, at least $C^{1}$) at the sonic point, and the family
of transonic solutions given in the subsonic region between the regular
center $z=0^{-}$ and the sonic point $z=z_{s}$ contains one parameter $D$ in
addition to the parameter $\alpha$.
The classification of the transonic solutions as well
as the clear presentation of the global spacetime structure has been
done \cite{OriA:PRD42:1990, FoglizzoT:PRD48:1993}.

The banded structure of the parameter $D$ is a notable point of such a
classification by the total number $N$ of
recurrent bounces and recollapses which may occur subsequently to an
initial collapse. 
The allowed range of $D$ for the $N$-th class of solutions is
limited to a finite band $D_{2N}\leq D \leq D_{2N+1}$ ($N=0, 1, 2,\cdots$).
If the solutions are required to be analytic (namely, $C^{\infty}$) at
the sonic point, the parameter $D$ is further restricted to some
discrete values lying in the $N$-th band.
For example, there exist only two no-bounce solutions (corresponding to
$N=0$) analytic at the sonic point \cite{OriA:PRD42:1990}.
They are called the flat Friedmann
solution and the general relativistic Larson-Penston solution, which
describe a homogeneous collapse and an inhomogeneous collapse,
respectively.
It is also remarkable that the perfect fluid critical solution
\cite{CarrBJ:PRD61:2000, NeilsenDW:CQG17:2000} corresponding to the threshold of black hole formation is the $N=1$
solution analytic at the sonic point.

Several works have been devoted to numerical analysis of the stability
problem for the self-similar perfect fluid solutions analytic at the
sonic point. In particular, it was numerically confirmed for the
parameter range $0<\alpha\lesssim 0.036$ that the flat Friedmann solution
and the general relativistic Larson-Penston solution can act as an
attractor in generic gravitational collapse \cite{HaradaT:PRD63:2001}, while the critical solution
for the parameter range $0<\alpha\lesssim 0.89 $ was numerically shown
to be unstable \cite{KoikeT:PRL74:1995, MaisonD:PLB366:1996,
KoikeT:PRD59:1999}.
(It is interesting to note that the general relativistic Larson-Penston
solution for the parameter range $0<\alpha\lesssim 0.0105$ describes naked
singularity formation \cite{OriA:PRD42:1990}, and thus it is a serious counterexample against the cosmic
censorship.)
Some of the $N\geq 2$ solutions were also numerically found to be unstable for
$0<\alpha\lesssim 0.036$ \cite{HaradaT:PRD63:2001} and $\alpha=1/3$ \cite{KoikeT:PRD59:1999}.
These numerical results obtained for limited values of $\alpha$ and $D$ suggest
that only the $N=0$ solutions can remain stable.
Such an interrelation between the stability and the
occurrence of bounces is also an interesting issue to be further
examined in the light of the analytical treatment of the stability
problem for any $\alpha$ and any allowed range of $D$.

We start from a brief review of general spherically symmetric perfect
fluid system in Sec.~\ref{sec:sp} and mention some key properties of the
self-similar solutions in Sec.~\ref{sec:background}.
Further, in Sec.~\ref{sec:waveeq}, the set of several linear equations for
non-self-similar matter and metric perturbations is derived, and the
perturbation equations are reduced to a single second-order wave
equation.
This construction of the single wave equation governing the matter and
metric perturbations will allow us to study analytically time evolution
of perturbations in a well-developed standard manner, for instance,
through the Green's function technique.
As a preliminary application of this master wave equation to the
stability problem, in Sec.~\ref{sec:normal}, we try to solve the eigenvalue problem of
the spectral parameter $\omega$ for perturbations having the time
dependence $\exp{(i\omega\log{|t|})}$ and satisfying the boundary
conditions (which exclude the kink modes studied in \cite{HaradaT:CQG18:2001}) at the regular
center $z=0^{-}$ and the sonic point $z=z_{s}$.
Then, as the main result in Sec.~\ref{sec:normal}, some sufficient
conditions for the absence of unstable modes with purely imaginary
$\omega$ in self-similar backgrounds (non-bouncing at least in the
subsonic region) are presented.
In the final section, we summarize the results obtained in this paper 
and discuss the consistency of our analytical scheme with the
numerical results obtained in previous studies.

\section{Spherically symmetric perfect fluid system\label{sec:sp}}
In this section we briefly review the general spherically symmetric perfect fluid
system as a first step to study matter and metric perturbations in
self-similar backgrounds, mainly referring to the paper \cite{HaradaT:PRD63:2001}.
The spherically symmetric line element is given by 
\begin{equation}
 ds^{2}=
  -e^{2\nu(t,r)}dt^{2}+e^{2\lambda(t,r)}dr^{2}+R^{2}(t,r)\left(d\theta^{2}+\sin^{2}\theta d\varphi^{2}\right)~,\label{eq:LE}
\end{equation}
with the comoving coordinates $t$ and $r$.
The collapsing matter is a perfect fluid whose energy momentum tensor is
given by
\begin{equation}
 T^{ab}=(\rho+P)u^{a}u^{b}+Pg^{ab}~,
\end{equation}
where $\rho$, $P$ and the vector $u^{a}$ are energy
density, pressure and fluid four velocity, respectively.
As was mentioned in Sec.~\ref{sec:intro}, we assume that its equation of state is 
\begin{equation}
 P = \alpha \rho~,
\end{equation}
using a constant $\alpha$ lying in the range $0<\alpha\leq 1$.
To discuss self-similar behaviors in the following, we use a new variable $z$
defined by $z\equiv r/t$, instead of $r$. 
In addition we also introduce the following dimensionless functions:
\begin{eqnarray}
 \eta (t,r) &\equiv& 8\pi r^{2}\rho~,\\
 S(t,r) & \equiv & \frac{R}{r}~.
\end{eqnarray}

From the Einstein's field equations,  
we can obtain the four equations governing the functions $\nu$, $\lambda$,
$S$ and $\eta$.
By virtue of the choice of the comoving coordinates, the two equations
lead to the relations 
\begin{eqnarray}
 e^{\nu}& =&  C_{\nu}(t)(z^{2})^{\alpha/(1+\alpha)}\eta^{-\alpha/(1+\alpha)}~\label{eq:nu},\\
 e^{\lambda}&=&C_{\lambda}(r)\eta^{-1/(1+\alpha)}S^{-2}~\label{eq:lambda},
\end{eqnarray}
where $C_{\nu}$ and $C_{\lambda}$ are arbitrary functions.
Thus the remaining two equations become the equations for only the
two unknown functions
$S$ and $\eta$ and are written by
\begin{eqnarray}
 && M+M'=\eta S^{2}(S+S')~,\label{eq:E1b}\\
 && \dot{M}-M'=-\alpha\eta S^{2}(\dot{S}-S')~,\label{eq:E2b}
\end{eqnarray}
where the dot and the prime represent the partial derivative with
respect to $\log|t|$ and $\log|z|$, respectively.
The function $M$ introduced in Eqs.~(\ref{eq:E1b}) and (\ref{eq:E2b}) is defined by
\begin{equation}
 M (t,z)\equiv S\left\{1+e^{-2\nu}z^{2}\left(\dot{S}-S'\right)^{2}-e^{-2\lambda}\left(S+S'\right)^{2}\right\}~.\label{eq:M}
\end{equation} 
Note that the function $M$ is the dimensionless function related to the
Misner-Sharp mass $m$ as $M=2m/r$.
In the following Eqs.~(\ref{eq:E1b}) and (\ref{eq:E2b}) will act as the basic equations to
derive the perturbation equations.

Finally in this section we define another important quantity $V$ as 
\begin{equation}
 V(t,z) \equiv  ze^{\lambda-\nu}~. 
\end{equation}
This is interpreted as the velocity of a $z=\text{const}$
surface relative to the fluid element. 

\section{Background self-similar spacetimes\label{sec:background}}
Now we point out some features of spherically symmetric self-similar backgrounds. 
In particular we focus our concern on their asymptotic behaviors
 near the regular center and near the sonic point, at which some
 boundary conditions are imposed on the perturbations.
The self-similarity considered here means that 
\begin{equation}
 \nu=\nu(z)~, \qquad \lambda=\lambda(z)~, \qquad S=S(z)~, \qquad
  \eta=\eta(z)~, \label{eq:background1}
\end{equation}
which require $C_{\nu}$ and $C_{\lambda}$ be constant. 
Then,  Eqs.~(\ref{eq:E1b}),
(\ref{eq:E2b}) and (\ref{eq:M}) are reduced to the ordinary differential
equations, which are written as
\begin{eqnarray}
 && M+M'=\eta S^{2}\left(S+S'\right)~,\label{eq:E1c}\\
&& M'=-\alpha\eta S^{2}S'~,\label{eq:E2c}\\
&& M (z) = S\left\{1+e^{-2\nu}z^{2}S^{\prime 2}-e^{-2\lambda}\left(S+S'\right)^{2}\right\}~\label{eq:Mb}.
\end{eqnarray}

It can be easily found that the Ricci scalar constructed by the self-similar metrics indefinitely
increases as $t$ approaches zero along a $z=\text{const}$ line. 
Thus, in the limit $t\rightarrow 0$, a singularity will appear at the center $r=0$ in the self-similar spacetime. 
Hereafter we are interested in the time evolution of the system before
the singularity formation, i.e., $t<0$ (i.e., $z<0$).

Because the derivative of the function $S$ along a $r=\text{const}$ line
leads to 
\begin{equation}
 z\frac{dS}{dz}=-t\frac{dS}{dt},
\end{equation}
the inequality $S'>0$ (or $S'<0$) means a local expansion (or local
contraction) of a fluid shell.
Taking account of such a physical meaning of the function $S'$, we
here introduce our original expression for Eqs.~(\ref{eq:E1c}),
(\ref{eq:E2c}) and (\ref{eq:Mb}) which is useful for seeing the motion
of fluid; we rewrite Eqs.~(\ref{eq:E1c}),
(\ref{eq:E2c}) and (\ref{eq:Mb}) in terms of the
function $S'$ (or $A'\equiv S'/S$) and the velocity $V$, instead of the functions $\eta$ and $S$.
From Eqs.~(\ref{eq:nu}) and (\ref{eq:lambda}) we find that the function
$V$ is related with the functions $\eta$ and $S$ by
\begin{equation}
 V= -C_{\lambda}C_{\nu}^{-1}(-z)^{(1-\alpha)/(1+\alpha)}\eta^{(\alpha-1)/(1+\alpha)}S^{-2}~.\label{eq:Vb}
\end{equation}
In addition the function $M$ can be written in terms of the functions $A'$ and
$V$ as
\begin{equation}
 M=S\left[1+X\left\{V^{2}A^{\prime2}-\left(1+A'\right)^{2}\right\}\right]~,\label{eq:Mc}
\end{equation}
where the function $X$ is defined by
\begin{equation}
 X \equiv S^{2}e^{-2\lambda}~.
\end{equation}
Because of Eq.~(\ref{eq:lambda}), 
the function $X$ is also rewritten as 
\begin{equation}
 X = C_{\lambda}^{-2}\eta^{2/(1+\alpha)}S^{6}~.\label{eq:Xb}
\end{equation}
Using Eqs.~(\ref{eq:Vb}), (\ref{eq:Mc}) and (\ref{eq:Xb}),
we can reduce Eqs.~(\ref{eq:E1c}), (\ref{eq:E2c}) and (\ref{eq:Mb}) to
\begin{eqnarray}
 &&\frac{V'}{V}
=\frac{(1-\alpha^{2})A''+(1+\alpha)(1-5\alpha)A^{\prime2}-2(\alpha^{2}+2\alpha-1)A'+1-\alpha}{(1+\alpha)\left\{1+(1+\alpha)A'\right\}}~,\label{eq:Veq}\\
 &&\frac{X'}{X}=\frac{-2(1+\alpha)A''+6\alpha(1+\alpha)A^{\prime2}+4\alpha
  A'}{(1+\alpha)\left\{1+(1+\alpha)A'\right\}}~,\label{eq:Xeq}\\
&&2X(1+\alpha)\left(V^{2}-\alpha\right)A''+(1-\alpha)XV^{2}A'\left\{3(1+\alpha)A'+2\right\}\nonumber\\
&&\hspace{1.5cm}-X\left(1+A'\right)\left\{(1+\alpha)(1+7\alpha)A'+\alpha^{2}+6\alpha+1\right\}+(1+\alpha)^{2}=0~.\label{eq:A''}
\end{eqnarray}
Note that the use of the functions $X$ and $A'$ instead of $M$ and $S'$ prevents the functions
$S$ and $\eta$ and the variable $z$ from explicitly appearing in
Eqs.~(\ref{eq:Veq}), (\ref{eq:Xeq}) and (\ref{eq:A''}) to reduce
the field equations to a simpler autonomous system.
In this paper we will mainly represent the self-similar solutions by the
functions $V$, $A'$ and $X$.

One of the requirements for the solutions to describe gravitational
collapse beginning with the regular initial data is their regularity at
the center $r=0$ during $t<0$ (i.e., $z=0^{-}$). 
The approximate behaviors of the solutions $S$, $\eta$ and $M$ satisfying
the regularity condition near the center $z=0^{-}$ are given in
\cite{HaradaT:PRD63:2001}, from which the behavior of the function $V$
near the regular center is found to be 
\begin{equation}
 V \simeq -(1-p)^{2/3}C_{\lambda}^{1/3}(2D)^{-p/2}(-z)^{1-p}~,\label{eq:Vc}
\end{equation}
where $p$ and $D$ are constants defined as 
\begin{eqnarray}
 p &\equiv& \frac{2}{3(1+\alpha)}~,\label{eq:p}\\
 D &\equiv& 4\pi\rho(t,0)t^{2}~.
\end{eqnarray}
In addition we can also find that at the regular center
\begin{equation}
 A' = -p~, \qquad X = (1-p)^{-2}.\label{eq:A'c}
\end{equation}
Because the constant $C_{\lambda}$ is just the freedom of rescaling the
radial coordinate, Eq.~(\ref{eq:Vc}) means that the self-similar
solutions with the regular center are characterized by one parameter $D$
for a given $\alpha$.  

From Eq.~(\ref{eq:A''}) we see that the self-similar perfect fluid system
may become singular when the velocity of a $z=\text{const}$
surface relative to the fluid is equal to the sound speed, i.e.,
$V^{2}=\alpha$. We call this singular point in the equation a sonic
point and denote its location by
$z=z_{s}$. 
For convenience we hereafter denote the value of a function at the
sonic point by attaching the suffix ``$S$'' to the function (e.g.,
$A'(z_{s})=A'_{s}$).

Because of Eq.~(\ref{eq:Vc}), the function $V$ becomes zero at
$z=0^{-}$.
This means that the flow speed is subsonic at $t=-\infty$.
However, with the lapse of time, the flow speed of a collapsing shell
with a constant $r$ will get larger.
Thus in gravitational collapse
the velocity $V$ (or $A'$ and $X$) can smoothly decrease to a supersonic
value $V<-\sqrt{\alpha}$ beyond the sonic point $z=z_{s}$ as $z$
decreases.
The asymptotic behaviors of the self-similar solutions near the sonic point and the transonic conditions
were examined in e.g., \cite{BicknellGV:AJ225:1978} and
\cite{OriA:PRD42:1990}.
We here briefly review such studies, using the result obtained in
Appendix~\ref{sec:sonic}, in which the asymptotic behaviors of our
original set of solutions $V$, $A'$ and $X$ near the sonic point are examined.

The transonic solutions, namely, the solutions which are at least smooth
at the sonic point, are well parametrized by $A'_{s}$ (or $V'_{s}$) 
in the region near the sonic point. 
Note that $V'_{s}$ is necessarily negative.
In addition the terms
proportional to $(\xi-\xi_{s})^{\chi}$ with non-integer power index $\chi$
generally appear in the expansion of the solutions near the sonic point, 
where $\xi$ is defined by $\xi\equiv \log(-z)$, and the power index
$\chi$ is given by 
\begin{equation}
 \chi = -1-\frac{\sqrt{\alpha}}{V'_{s}}.\label{eq:chis}
\end{equation}
In order for the solutions to be smooth at the sonic point, the power index $\chi$ must be larger than unity, otherwise the
coefficients ($V_{\chi}$, $A'_{\chi}$ and $X_{\chi}$ written in
Appendix~\ref{sec:sonic}) of the terms including
$(\xi-\xi_{s})^{\chi}$ must vanish by a fine tuning.
For the former condition the solutions are smooth but not analytic at
the sonic point, while the solutions are analytic at the sonic point for the
latter condition.

From Eq.~(\ref{eq:chis}) we see that the transonic condition $\chi\geq 1$ restrict the allowed range of $V'_{s}$.
Because for a given $\alpha$ the value of $V'_{s}$ and the coefficients
of the terms including $(\xi-\xi_{s})^{\chi}$ must depend on the
parameter $D$, 
the transonic conditions seem to also restrict the range of possible
values of $D$.
In fact, as was mentioned in Sec.~\ref{sec:intro}, the banded structure
$D_{2N}\leq D \leq D_{2N+1}$ ($N=0, 1, 2,\cdots$) of the allowed range
of $D$ for the
transonic solutions and the discretization
of the allowed values of $D$ for the solutions analytic at
the sonic point were numerically found.
One of such discrete values of $D$ is $D_{F}\equiv 2/3(1+\alpha)^{2}$, for
which we can find the solution for
Eqs.~(\ref{eq:Veq}), (\ref{eq:Xeq}) and (\ref{eq:A''}) written by 
\begin{equation}
 V=V_{F}(-z)^{1-p}~, \qquad A' = -p~, \qquad X=(1-p)^{2}\label{eq:FF}
\end{equation}
with the constant $V_{F}$ dependent only on $\alpha$, which is called
the flat Friedmann solution.
It should be noted that $D_{F}$ lies in the 0-th band, namely, in the range 
$D_{0}\leq D_{F} \leq D_{1}$.

As was also mentioned in Sec.~\ref{sec:intro}, it is remarkable that the number of zeros of
the solution $A'$, namely, the total number of turns of radial motion of a
fluid shell from a contraction to an expansion and from an expansion to
a contraction, for $D$ lying in the $N$-th band is equal to $N$.\footnote{In \cite{OriA:PRD42:1990} Ori and Piran originally revealed the relation between the number of the zeros of the radial velocity
$u^{r}$ of the fluid in the non-comoving coordinate system and the rank of the permitted band which the
solution belongs to. The velocity $u^{r}$ can be related to the
quantities in the comoving coordinate system used in this paper as
$u^{r}=VX^{1/2}A'$ (see \cite{OriA:PRD42:1990} for the coordinate
transformation from the non-comoving coordinate system to the comoving one). 
Because the functions $V$ and $X$ cannot be zero for $z<0$ by
definition, the zeros of the velocity $u^{r}$ correspond to those of the function $A'$. }
For example, as seen from Eq.~(\ref{eq:FF}), the function $A'$ for the flat
Friedmann solution does not become zero at any $z$, namely, the flat
Friedmann solution describes collapse with no-bounce.

\section{Master wave equation governing perturbations\label{sec:waveeq}}

Now we consider spherically symmetric non-self-similar perturbations in
self-similar backgrounds describing the gravitational collapse of a perfect fluid. 
Let us begin with expressing the solutions of the field equations (\ref{eq:E1b}) and 
(\ref{eq:E2b}) as
\begin{eqnarray}
 &&S(t,z)=S_{B}(z)\left\{1+\epsilon S_{1}(t,z)+O(\epsilon^{2})\right\}~,
  \qquad
  \eta(t,z)=\eta_{B}(z)\left\{1+\epsilon\eta_{1}(t,z)+O(\epsilon^{2})\right\}~,\nonumber\\ 
 &&M(t,z)=M_{B}(z)\left\{1+\epsilon M_{1}(t,z)+O(\epsilon^{2})\right\}
\end{eqnarray}
with a small parameter $\epsilon$. 
We use the freedom of the coordinate transformation for 
$t$ and $r$ to require the constants $C_{\nu}$ and $C_{\lambda}$ in
Eqs.~(\ref{eq:nu}) and (\ref{eq:lambda}) be non-perturbed, namely, 
\begin{equation}
 C_{\nu} = C_{\nu B}~, \qquad C_{\lambda}=C_{\lambda B}~.
\end{equation}
Then the perturbations for the metrics $\nu$ and $\lambda$ are written
by $S_{1}$ and $\eta_{1}$ via Eqs.~(\ref{eq:nu}) and (\ref{eq:lambda}).
In addition we can easily write the perturbation
$\eta_{1}$ by $S_{1}$ and $M_{1}$, using the
perturbation equations (up to the linear order of $\epsilon$) 
derived from Eqs.~(\ref{eq:E1b}), (\ref{eq:E2b}) and
(\ref{eq:M}) which do not contain the derivative of $\eta$.
We thus obtain the following two first-order partial differential equations for $S_{1}$ and $M_{1}$:
\begin{eqnarray}
 &&P_{1}(z)M'_{1}+P_{2}(z)M_{1}+P_{3}(z)\dot{S}_{1}+P_{4}(z)S'_{1}+P_{5}(z)S_{1}=0,\label{eq:P}\\
 &&P_{1}(z)\dot{M}_{1}-P_{1}(z)M'_{1}+Q_{2}(z)M_{1}+Q_{3}(z)\dot{S}_{1}+Q_{4}(z)S'_{1}+Q_{5}(z)S_{1}=0
  \label{eq:Q}
\end{eqnarray}
with the coefficients $P_{i}$ and $Q_{i}$ written by the
background self-similar solutions, which are given in Appendix~\ref{sec:functions}.
In this paper we focus our analysis on the perturbations $S_{1}$ and
$M_{1}$. 
Hereafter we omit the suffix ``$B$'' attached to the background
self-similar solutions for simplicity.

The complicated form of Eqs.~(\ref{eq:P}) and (\ref{eq:Q}) may enforce
us to study numerically time evolution of the perturbations.
Nevertheless, we can expect that it becomes easier to understand
analytically its essential features if the two equations are reduced to
a single second-order wave equation.
Here we adopt such a strategy for our perturbation analysis.
For this purpose we introduce the function $\Psi$ written as 
\begin{equation}
 \Psi(t,z) =S_{1}(t,z)-f(z)M_{1}(t,z)
\end{equation}
with an arbitrary function $f(z)$.
Then, from Eqs.~(\ref{eq:P}) and (\ref{eq:Q}) we can obtain the two first-order partial differential equations
for $\Psi$ and $M_{1}$ with the coefficients containing the
function $f$.
It is easy to see that if the function $f$ is given by
\begin{equation}
 f= f_{\pm}\equiv \frac{\left\{1+(1+\alpha)A'\right\}\left\{\sqrt{\alpha}VA'\mp\left(1+A'\right)\right\}}{\sqrt{\alpha}\left\{VA'\pm\sqrt{\alpha}\left(1+A'\right)\right\}}~,\label{eq:f}
\end{equation}
from the equations for $\Psi$ and $M_{1}$ we obtain the relation in
which both $M'$ and $\dot{M}$ are eliminated as follows, 
\begin{equation}
 Y(z)M_{1}=T_{1}(z)\dot{\Psi}+T_{2}(z)\Psi'+T_{3}(z)\Psi,\label{eq:T}
\end{equation}
where the coefficients $Y$ and $T_{i}$ are written by the functions $P_{i}$,
$Q_{i}$ and $f$ (see Appendix~\ref{sec:functions}).
By using Eq.~(\ref{eq:T}) to eliminate $M_{1}$ from the equations for
$\Psi$ and $M_{1}$, we arrive at the following single equation for
$\Psi$ only:
\begin{equation}
 \ddot{\Psi}-2\dot{\Psi}'+\left(1-\frac{\alpha}{V^{2}}\right)\Psi''+R_{1}(z)\dot{\Psi}+R_{2}(z)\Psi'+R_{3}(z)\Psi=0,\label{eq:wave}
\end{equation}
where the coefficients are also written by the functions
$P_{i}$ and $Q_{i}$ and $f$. 
As will be shown in Sec.~\ref{sec:gauge}, a gauge mode corresponding to
an infinitesimal change of the background self-similar solution due to a
transformation of the time coordinate $t$ is allowed as a solution of
Eq.~(\ref{eq:wave}).
Because no other choice of $f$ different from Eq.~(\ref{eq:f}) is
possible for obtaining the second-order wave equation (\ref{eq:wave}),
our scheme is not equivalent to a so-called gauge-invariant approach for
perturbations.

Here we consider the transformations of the variables in
Eq.~(\ref{eq:wave}), which are given by
$t\rightarrow u=\log(-t)+I(\xi)$ and $\xi\rightarrow \zeta(\xi)$, 
where the functions $I$ and $\zeta$ are defined as
\begin{equation}
 I(\xi) =
  \int^{\xi}_{-\infty}\frac{V^{2}d\xi}{V^{2}-\alpha}~,\qquad \zeta(\xi) = \int^{\xi}_{-\infty}\frac{\sqrt{\alpha}Vd\xi}{V^{2}-\alpha}~.\label{eq:zeta}
\end{equation}
The new spacial variable $\zeta$ runs from zero to positive infinity in
the region between the regular center
$\xi\rightarrow -\infty$ and the sonic point $\xi\rightarrow\xi_{s}$.
By virtue of the transformation (\ref{eq:zeta}), Eq.~(\ref{eq:wave}) is
reduced to the standard form of the wave equation:
\begin{equation}
 \Psi_{,uu}-\Psi_{,\zeta\zeta}+W(\zeta)\Psi_{,u}+F(\zeta)\Psi_{,\zeta}+U(\zeta)\Psi=0~.\label{eq:wave2}
\end{equation}
We can explicitly write the functions $W$, $F$ and $U$ by the
background self-similar solutions $V$, $A'$ and $X$, which are shown in 
Appendix~\ref{sec:functions}. 

Though in general we cannot see the detailed dependence of the functions $W$, $F$
and $U$ on 
the variable $\zeta$ without numerically
solving the field equations (\ref{eq:Veq}), (\ref{eq:Xeq}) and
(\ref{eq:A''}), we can analytically find their general behaviors near the
boundaries. 
In the following calculations we choose the function $f$ as $f_{+}$ 
because by virtue of this choice the behaviors of the functions $W$, $F$
and $U$ near the sonic point (at which $V=-\sqrt{\alpha}$) become much simpler.
Because of Eq.~(\ref{eq:Vc}) and the definition of the variable $\zeta$,
the asymptotic behavior of the function $V$ near the regular center in terms of the variable
$\zeta$ is
\begin{equation}
 V \simeq \frac{\sqrt{\alpha}(3\alpha+1)}{3(1+\alpha)}\zeta~.
\end{equation}
Using this relation, we find that the functions $W$, $F$ and $U$ become
infinitely large near the regular center as
\begin{eqnarray}
 W &\simeq &\frac{1}{\zeta}~,\label{eq:Wc}\\
 F & \simeq &  -\frac{3}{\zeta}~,\label{eq:Fc}\\
 U &\simeq & \frac{3}{\zeta^{2}}~,\label{eq:Uc}
\end{eqnarray}
without depending on the parameters $D$ and $\alpha$.
In addition, from Eqs.~(\ref{eq:X0}), (\ref{eq:A'1}) and (\ref{eq:V1}), 
we can find that the values $W_{s}$, $F_{s}$ and $U_{s}$ of the
functions $W$, $F$ and $U$  at the sonic point are given by
\begin{eqnarray}
 W_{s}&=&-F_{s}=2\left(1-\frac{V'_{s}}{\sqrt{\alpha}}\right)~,\label{eq:Ws}\\
 U_{s}&=&0~.\label{eq:Us}
\end{eqnarray}
Taking account of such behaviors of the functions $W$, $F$ and $U$ in
Eq.~(\ref{eq:wave2}) near the
boundaries, we will consider the boundary
conditions for $\Psi$ in the next section.

In addition we should mention the behaviors of the functions $W$, $F$
and $U$ near a point at which $A'=0$, namely, radial motion of matter stops.
Let us denote this point by $\zeta=\zeta_{0}$.
The functions $W$, $F$ and $U$ apparently diverge at $\zeta=\zeta_{0}$
in proportion to $(\zeta-\zeta_{0})^{-1}$. 
In particular, by using the field equation (\ref{eq:A''}) to see the
leading term of $A'$ near $\zeta=\zeta_{0}$, 
we find that the leading form of the function $F$ near $\zeta=\zeta_{0}$ is
\begin{equation}
 F \simeq \frac{1}{\zeta-\zeta_{0}}~.
\end{equation}
Then, it is easy to check that the solution $\Psi$ of
Eq.~(\ref{eq:wave2}) should remain finite or vanish in proportion to
$(\zeta-\zeta_{0})^{2}$ in the limit $\zeta\rightarrow\zeta_{0}$.
Therefore the wave equation (\ref{eq:wave2}) is applicable even to a background self-similar solution for the parameter $D$ lying in the
permitted band labelled by $N\geq 1$, namely, for a class in which
at least one bounce of a fluid shell in gravitational collapse is allowed.
 
\begin{figure}
\includegraphics[width=8cm]{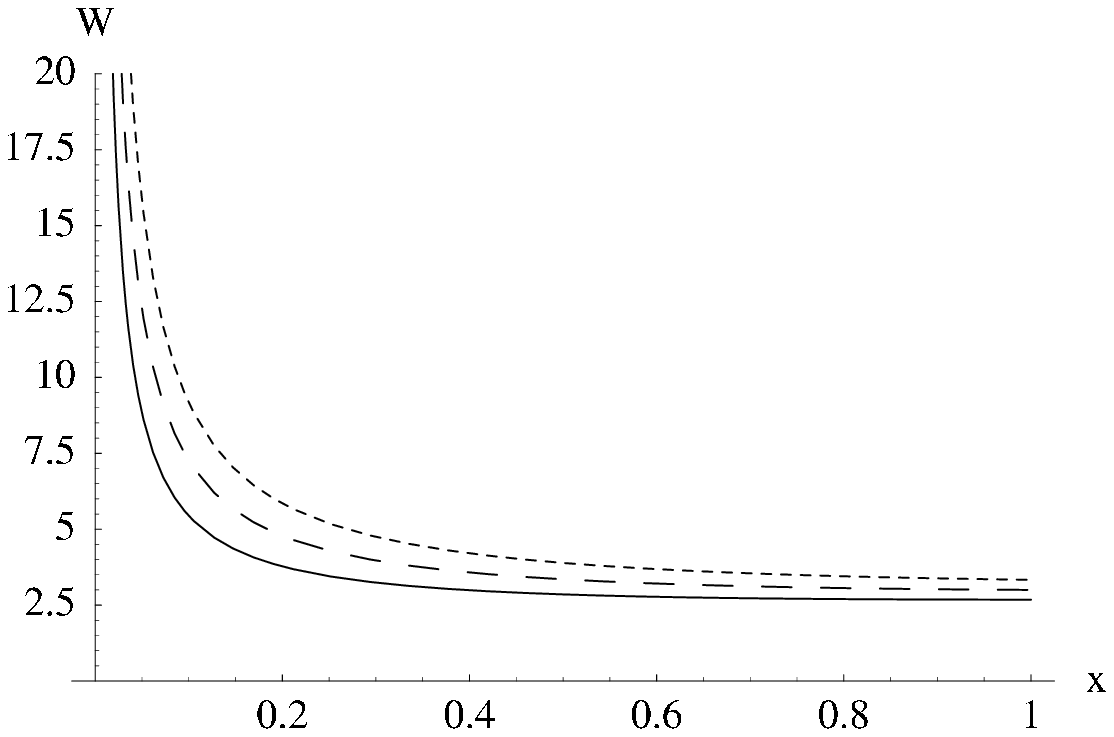}
\includegraphics[width=8cm]{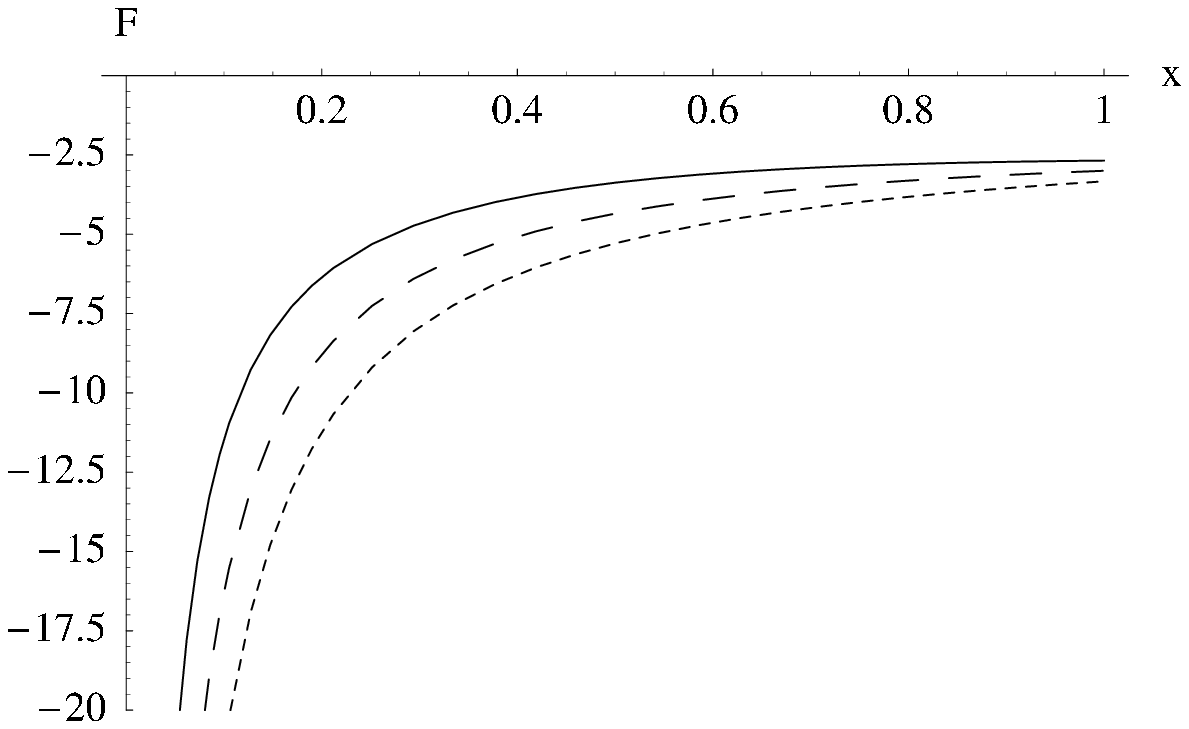}
\includegraphics[width=8cm]{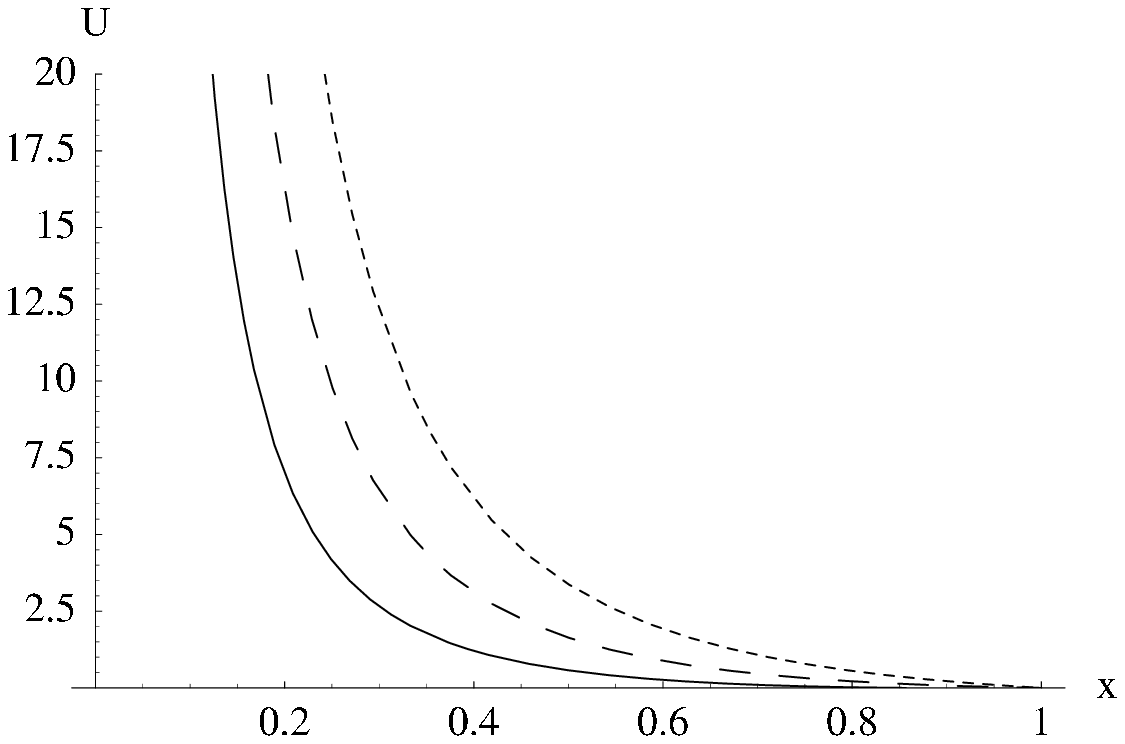}
\caption{Variations of $W$, $F$ and $U$ for the
 flat Friedmann solution corresponding to
 $\alpha=0.01$ (solid line), $\alpha=1/3$ (broken line) and $\alpha=1$
 (dotted line).}
\label{fg:FF}
\end{figure}

In the case that the background self-similar solution is the flat Friedmann
solution, which is given by Eq.~(\ref{eq:FF}), we can explicitly write the functions $W$, $F$ and $U$ as functions of $z$.
In Fig.~\ref{fg:FF} we draw their variations as functions of $x$ defined by
\begin{equation}
 x=-\frac{V}{\sqrt{\alpha}}~,\label{eq:x}
\end{equation}
which becomes zero at the regular center and unity at
the sonic point.
The functions $W$ and $-F$ are monotonic and always positive for any
$\alpha$ in the range $0<\alpha\leq 1$.
In addition the function $U$ is monotonic and always positive for $1/9\leq\alpha\leq1$, 
but not for $0<\alpha< 1/9$, which will be understood by
Eqs.~(\ref{eq:Uc}), (\ref{eq:Us}) and the derivative
\begin{equation}
 U_{,x}(1)=\frac{2(1+3\alpha)(1-9\alpha)}{9(1+\alpha)^{2}}~,
\end{equation}
though the range of $x$ for which $U<0$ is too small to be seen
in Fig.~\ref{fg:FF}.
We expect that such variations of the
functions $W$, $F$ and $U$ shown in Fig.~\ref{fg:FF} may be typical for 
self-similar solutions for the parameter $D$ lying in the $0$-th
band, namely, for solutions describing collapse with no-bounce.

It should be emphasized that the wave equation (\ref{eq:wave2}) is applicable to spherical
perturbations in any spherically symmetric self-similar perfect fluid
solutions (in particular, solutions which is smooth but may not be analytic at the sonic
point) describing the gravitational collapse.
Therefore we expect the wave equation to be useful 
for revealing time evolution of the perturbations for various types of
models of self-similar collapse covering a wide range of the parameter values $D$ and
$\alpha$.
As an important application of the wave equation (\ref{eq:wave2}), we will consider the stability problem for
self-similar collapse in the next section.

\section{Normal mode analysis\label{sec:normal}}

As was mentioned in Sec.~\ref{sec:intro}, the stability of self-similar
backgrounds for the perturbations has been numerically studied in many
works.
Here we would like to use the wave equation (\ref{eq:wave2}) for an
analytical approach to the stability problem.
Though our analysis done in this section remains preliminary for future
investigations, some clear proofs supporting the numerical results will
be successfully presented.

\subsection{boundary conditions}
By virtue of the self-similarity of the background, (namely, the fact
that the functions $W$, $F$ and $U$ depend on $z$ only)
we can reduce the wave equation (\ref{eq:wave2}) to the ordinary
differential equation as follows, 
\begin{equation}
 \psi_{,\zeta\zeta}-F\psi_{,\zeta}+\left(\omega^{2}-i\omega W-U\right)\psi=0~,\label{eq:wave3}
\end{equation}
by assuming that the wave function $\Psi$ is written as
\begin{equation}
 \Psi(u,\zeta) = \psi(\zeta,\omega)e^{i\omega u}.\label{eq:Psiu}
\end{equation}
Owing to the behaviors of $W$, $F$ and $U$ mentioned in the previous section, the general solution (or its higher
derivative) of Eq.~(\ref{eq:wave3}) may become singular at the
regular center or at the sonic point.
Therefore we study the conditions which the function $\psi$ should satisfy
at the boundaries. 

From Eqs.~(\ref{eq:Wc}), (\ref{eq:Fc}) and
(\ref{eq:Uc}), we obtain the asymptotic behavior of the general solution of
Eq.~(\ref{eq:wave3}) near the regular center as 
\begin{equation}
 \psi \simeq C_{1}(\omega)\zeta + C_{2}(\omega)\zeta^{-3}~.
\end{equation}
The divergence of $\psi$ implicates that of $M_{1}$ because of
Eq.~(\ref{eq:T}).
Thus we require that $\psi$ becomes
\begin{equation}
 \psi \sim \zeta~,\label{eq:psic}
\end{equation}
near the center.
Note that this boundary condition at the regular center is valid for
any self-similar backgrounds.

Because of Eqs.~(\ref{eq:Ws}) and (\ref{eq:Us}), near
the sonic point the general solution of Eq.~(\ref{eq:wave3}) behaves as
\begin{equation}
 \psi \simeq C_{3}(\omega)e^{i\omega\zeta}+C_{4}(\omega)e^{-\left(W_{s}+i\omega\right)\zeta}~.\label{eq:psis}
\end{equation}
Because near the sonic point we have 
\begin{equation}
 I \simeq \zeta~,
\end{equation}
Eq.~(\ref{eq:psis}) leads to 
\begin{equation}
 \Psi \simeq \left\{C_{3}(\omega)+C_{4}(\omega)e^{-\left(W_{s}+2i\omega\right)\zeta}\right\}e^{i\omega\log(-t)}~,\label{eq:Psib}
\end{equation}
from which we note that unstable modes (i.e., modes which grow as $t\rightarrow 0$) correspond to solutions $\Psi$ for $\omega$ whose
imaginary part is positive.
The boundary condition imposed on $\Psi$ at the sonic point may be a
subtle problem.
To see this, let us consider the leading behavior of perturbed physical
quantities (for instance, $\eta_{1}$) near the sonic point.
Using Eq.~(\ref{eq:T}) and the perturbation equation obtained from
Eq.~(\ref{eq:E1b}), we find that $\eta_{1}$ contains the second
derivative $\Psi''$.
Then, by taking account of Eqs.~(\ref{eq:Psib}), (\ref{eq:chis}) and (\ref{eq:Ws}) we
can express the leading behavior of $\eta_{1}$ as 
\begin{equation}
 \eta_{1}(t,z) \simeq \left\{\eta_{1a}(\omega)+\eta_{1b}(\omega)\left(\xi-\xi_{s}\right)^{\chi-(\sqrt{\alpha}i\omega/V'_{s})}\right\}e^{i\omega\log(-t)}~,\label{eq:eta1}
\end{equation}
which should be compared with the background solution $\eta_{B}$.
If $\eta_{B}$ is analytic at the sonic point, it is plausible that the
perturbation $\eta_{1}$ is also required to be analytic there, that is,
$\eta_{1b}$ is required to be zero.
However, the background solution may be regular (i.e., $C^{1}$), but not
be analytic there.
Then, the non-analytic term in proportion to $(\xi-\xi_{s})^{\chi}$
appears in $\eta_{B}$.
This fact and Eq.~(\ref{eq:eta1}) mean that for unstable modes (i.e., modes with
$\text{Im}(\omega)>0$), the perturbation becomes less regular than the
background quantity because $V'_{s}$ is negative.
One may claim that the higher derivatives of $\eta_{1}$ must remain
smaller than the corresponding higher derivatives of $\eta_{B}$ (like 
$\eta_{1}$ for $\eta_{B}$ analytic at the sonic point).
Then, $\eta_{1b}$ in Eq.~(\ref{eq:eta1}) should be zero even if the
background is not analytic at the sonic point, and the solution for
Eq.~(\ref{eq:wave3}) must be derived under the condition that the
leading behavior of $\psi$ near the sonic point is given by 
\begin{equation}
 \psi \sim e^{i\omega\zeta}~.\label{eq:psins}
\end{equation}
Though whether the boundary condition (\ref{eq:psins}) is physically
necessary is not so conclusive, hereafter we use it as well as
the boundary condition Eq.~(\ref{eq:psic}) to set up the eigenvalue
problem for Eq.~(\ref{eq:wave3}) and to obtain the normal modes and the
discrete eigenvalues denoted by $\psi_{n}$ and $\omega_{n}$.

\subsection{gauge mode \label{sec:gauge}}

We can find an exact solution $\psi_{g}$ for Eq.~(\ref{eq:wave3})
written by
\begin{equation}
 \psi_{g} =
  \frac{c(1+\alpha)VA^{\prime2}}{VA'+\sqrt{\alpha}(1+A')}e^{-i\omega_{g}I},\label{eq:psig}
\end{equation}
if the spectral parameter $\omega$ is equal to 
\begin{equation}
 \omega = \omega_{g} \equiv \frac{1-\alpha}{1+\alpha}i~,\label{eq:omegag}
\end{equation}
where $c$ is an arbitrary constant.
We can easily find that the solution $\psi_{g}$ satisfies the boundary
conditions.
Hence $\omega_{g}$ is one of the eigenvalues.
This normal mode corresponds to the correction of $S_{B}$ and $M_{B}$ generated by the following
infinitesimal coordinate transformation:
\begin{equation}
 (-t) \rightarrow (-t)\left\{1+\epsilon c(-t)^{i\omega_{g}}\right\}~, \qquad
  r \rightarrow r~,
\end{equation}
which means 
\begin{equation}
 (-z) \rightarrow (-z)\left\{1-\epsilon c(-t)^{i\omega_{g}}\right\}~.
\end{equation}
Therefore we call the solution $\psi_{g}$ a gauge mode.
Though the gauge mode is obviously unphysical, 
the presence of the solution explicitly written by the background
solutions $V$ and $A'$ will become mathematically useful for analysing
Eq.~(\ref{eq:wave3}), as will be seen later.

\subsection{Constraints for unstable modes}

In this subsection we present some analyses of Eq.~(\ref{eq:wave3}),
which is slightly different from the well-studied Strum-Liouville type
differential equation.
Because we are particularly interested in the stability of the
self-similar solutions, we examine the possibility of the existence of
unstable normal modes.
Hereafter we focus our investigation on the eigenvalues for the
background solutions in which the function $A'$ does not become zero at least in
the range between the regular center and the sonic point, to avoid the mathematical complexity caused
by the singular behavior of the functions $W$, $F$ and $U$ at $A'=0$.
Such solutions include all the solutions for the parameter $D$ lying in
the 0-th band, namely, for the class in which no bouncing motion of a
fluid shell occurs in collapse.
In addition the critical solution for $0.61 \lesssim \alpha \lesssim 1$ is also
included, because the zero of the function $A'$ is only in the supersonic
region \cite{CarrBJ:CQG18:2001}.
Therefore, the analysis done in this subsection may be applicable to
various types of solutions describing collapse accompanied with some bounces and recollapses.

Now let us begin with the introduction of the function $\phi$ defined by 
\begin{equation}
 \phi(\zeta,\omega) = \psi e^{-i\omega\zeta} \label{eq:phi}
\end{equation}
instead of $\psi$, because normal modes denoted by 
$\phi_{n}=\psi_{n}\exp{(-i\omega_{n}\zeta)}$ become constant at the
sonic point.
In addition it will be mathematically convenient to use the variable
$x$ instead of $\zeta$ in the following calculations, which can cover
the whole region between the
regular center and the sonic point in the finite range $0\leq x \leq 1$,
irrespective of the parameter values $\alpha$ and $D$.
Using the function $\phi_{n}$ and
the variable $x$, we rewrite Eq.~(\ref{eq:wave3}) as
\begin{eqnarray}
 &&\phi_{n,xx}-\frac{\sqrt{\alpha}x}{V'(1-x^{2})}\left\{2i\omega_{n}-F+\frac{V'(x^{2}+1)}{\sqrt{\alpha}x^{2}}+\frac{V''(1-x^{2})}{V'x}\right\}\phi_{n,x}\nonumber\\
&&\hspace{5cm}-\frac{\alpha
  x^{2}}{V^{\prime2}(1-x^{2})^{2}}\left\{i\omega_{n}(F+W)+U\right\}\phi_{n}=0~.\label{eq:wave4}
\end{eqnarray}

Moreover, denoting the real part and the imaginary part of
the eigenvalues $\omega_{n}$, respectively, by $\beta_{n}$ and $\gamma_{n}$, 
we rewrite Eq.~(\ref{eq:wave4}) to the following equation:
\begin{equation}
 \left(h\phi_{n,x}\right)_{,x}-\frac{2i\beta_{n}\sqrt{\alpha}x}{V'(1-x^{2})}h\phi_{n,x}-\frac{\alpha
  x^{2}}{V^{\prime2}(1-x^{2})^{2}}\left\{i\omega_{n}(F+W)+U\right\}h\phi_{n}=0~,\label{eq:wave5}
\end{equation}
where the function $h$ is given by
\begin{equation}
 \frac{h_{,x}}{h}=-\frac{\sqrt{\alpha}x}{V'(1-x^{2})}\left\{-2\gamma_{n}-F+\frac{V'(x^{2}+1)}{\sqrt{\alpha}x^{2}}+\frac{V''(1-x^{2})}{V'x}\right\}~,\label{eq:h}
\end{equation}
and can be always positive in the range $0\leq x \leq 1$.
Following the standard procedure, we consider to multiply
Eq.~(\ref{eq:wave5}) by the complex conjugate $\phi^{*}_{n}$ of
$\phi_{n}$ and obtain 
\begin{eqnarray}
 &&\phi^{*}_{n}\left(h\phi_{n,x}\right)_{,x}+\phi_{n}\left(h\phi^{*}_{n,x}\right)_{,x}-\frac{2i\beta_{n}\sqrt{\alpha}x}{V'(1-x^{2})}h\left(\phi^{*}_{n}\phi_{n,x}-\phi_{n}\phi^{*}_{n,x}\right)\nonumber\\
&&\hspace{3cm}-\frac{2\alpha
  x^{2}}{V^{\prime2}(1-x^{2})^{2}}\left\{U-\gamma_{n}(F+W)\right\}h|\phi_{n}|^{2}=0~.\label{eq:wave7}
\end{eqnarray}

Let us integrate Eq.~(\ref{eq:wave7}), taking account of the boundary condition such
that $\phi_{n}\sim x$ near the regular center $x=0$.
Because the asymptotic behavior of the function $h$ near the regular
center is
\begin{equation}
 h \sim x^{3}~,
\end{equation}
we have 
\begin{eqnarray}
 h\left(|\phi_{n}|^{2}\right)_{,x}&=&2\int^{x}_{0}h\left[\left|\phi_{n,x}\right|^{2}+\frac{\alpha
  x^{2}}{V^{\prime2}(1-x^{2})^{2}}\left\{U-\gamma_{n}(F+W)\right\}\left|\phi_{n}\right|^{2}\right]dx\nonumber\\
 &&\hspace{3cm}+2i\beta_{n}\int^{x}_{0}\frac{\sqrt{\alpha}x}{V'(1-x^{2})}h\left(\phi^{*}_{n}\phi_{n,x}-\phi_{n}\phi^{*}_{n,x}\right)dx~.\label{eq:wavex}
\end{eqnarray}
We now apply the boundary condition at the sonic point to
Eq.~(\ref{eq:wavex}).
Using Eq.~(\ref{eq:Ws}), we find that the asymptotic behavior of the
function $h$ near the sonic point is 
\begin{equation}
 h \sim (1-x)^{-a}~,\label{eq:hs}
\end{equation}
where $a$ is a constant defined as 
\begin{equation}
 a \equiv \frac{\sqrt{\alpha}}{V'_{s}}(\gamma_{n}-1)~.
\end{equation}
For $\gamma_{n}>1$ (i.e., $a<0$ because $V'_{s}$ is negative), the integrations in Eq.~(\ref{eq:wavex}) remain
finite at $x=1$, and the left hand side of Eq.~(\ref{eq:wavex}) vanishes
there, because $\phi_{n}$ and its derivative become constant there.
Thus we obtain the relation
\begin{eqnarray}
 &&\int^{1}_{0}h\left[\left|\phi_{n,x}\right|^{2}+\frac{\alpha
  x^{2}}{V^{\prime2}(1-x^{2})^{2}}\left\{U-\gamma_{n}(F+W)\right\}\left|\phi_{n}\right|^{2}\right]dx\nonumber\\
 &&\hspace{3cm}+i\beta_{n}\int^{1}_{0}\frac{\sqrt{\alpha}x}{V'(1-x^{2})}h\left(\phi_{n}^{*}\phi_{n,x}-\phi_{n}\phi^{*}_{n,x}\right)dx=0
  \label{eq:gamman}
\end{eqnarray}
for $\gamma_{n}>1$.
This clearly shows that if the inequality 
\begin{equation}
 U-\gamma_{n}(F+W)>0 \label{eq:ineq}
\end{equation}
holds for $\gamma_{n}>1$ and any $x$ in the range $0\leq x \leq 1$, then
there exists no normal mode with $\gamma_{n}>1$ and $\beta_{n}=0$.
It is noteworthy that for $\gamma_{n}>1$, Eq.~(\ref{eq:ineq}) is satisfied for the flat
Friedmann solution given by any $\alpha$ in the range $0<\alpha\leq1$.

Though we cannot derive a condition for the absence of unstable normal modes with
$\gamma_{n}>1$ and $\beta_{n}\neq 0$ from Eq.~(\ref{eq:gamman}), 
we can derive another constraint for normal modes $\phi_{n}$
with $\beta_{n}\neq 0$ from the multiplication of $\phi^{*}_{n}$ to
Eq.~(\ref{eq:wave5}) as follows,
\begin{equation}
 \phi^{*}_{n}\left(h\phi_{n,x}\right)_{,x}-\phi_{n}\left(h\phi^{*}_{n,x}\right)_{,x}-\frac{2i\beta_{n}\sqrt{\alpha}x}{V'(1-x^{2})}h\left(|\phi_{n}|^{2}\right)_{,x}-\frac{2i\beta_{n}\alpha
  x^{2}}{V^{\prime2}(1-x^{2})^{2}}(F+W)h|\phi_{n}|^{2}=0~.\label{eq:subtruction}
\end{equation}
For normal modes with $\gamma_{n}>1$, the integration of Eq.~(\ref{eq:subtruction}) gives 
\begin{equation}
 \int_{0}^{1}\frac{\sqrt{\alpha}x}{V'(1-x^{2})}h\left\{\left(|\phi_{n}|^{2}\right)_{,x}+\frac{\sqrt{\alpha}x}{V'(1-x^{2})}(F+W)|\phi_{n}|^{2}\right\}dx=0~.\label{eq:integral}
\end{equation}
Using Eq.~(\ref{eq:integral}), we can estimate the second term of the left
hand side of Eq.~(\ref{eq:gamman}).
Then, the condition (\ref{eq:gamman}) for normal modes with
$\gamma_{n}>1$ may be shown to be incompatible with the additional
condition (\ref{eq:integral}), for instance, if the background is the
flat Friedmann solution.
This is an interesting problem to be fully examined in future works.

Further, we note that if oscillatory unstable normal modes (i.e.,
unstable normal modes with $\beta_{n}\neq 0$) with $\gamma_{n}>W_{s}/2$
(i.e., $a<-1$) exist, Eq.~(\ref{eq:integral}) allows an estimation of
$\gamma_{n}$.
In fact, the partial integration of the term including
$(|\phi_{n}|^{2})_{,x}$ in Eq.~(\ref{eq:integral}) becomes possible for
$\gamma_{n}>W_{s}/2$, and we obtain
\begin{equation}
 \gamma_{n}=\frac{\displaystyle{\int_{0}^{1}\frac{\alpha
  x^{2}}{V^{\prime2}(1-x^{2})^{2}}Wh|\phi_{n}|^{2}dx}}{\displaystyle{2\int_{0}^{1}\frac{\alpha
  x^{2}}{V^{\prime2}(1-x^{2})^{2}}h|\phi_{n}|^{2}dx}}\equiv \frac{W_{\text{av}}}{2}~.
\end{equation}
This means that $\gamma_{n}$ should be equal to 
a half of an average $W_{\text{av}}$ of the function $W$.

We have presented Eq.~(\ref{eq:ineq}) as a sufficient condition
for the absence of non-oscillatory unstable modes (i.e., unstable modes
with $\beta_{n}=0$) with $\gamma_{n}>1$. Special attention to the case
$\beta_{n}=0$ will be meaningful because all the unstable normal modes
numerically found in previous works were non-oscillatory \cite{KoikeT:PRL74:1995, KoikeT:PRD59:1999,
HaradaT:PRD63:2001}.
Therefore the next task should be to analyze normal
modes with $\beta_{n}=0$ and $0<\gamma_{n}<1$.
It should be noted that there exists a gauge mode
$\phi_{g}=\psi_{g}\exp{(-i\omega_{g}\zeta)}$ as an unphysical normal
mode with $\beta_{n}=0$ and $0<\gamma_{n}<1$.
Therefore, to obtain the conditions for the physical normal modes
$\phi_{n}$, we must consider a difference between $\phi_{n}$ and
$\phi_{g}$.
For this purpose, we multiply Eq.~(\ref{eq:wave4}) for $\phi_{n}$ by $\phi_{g}$ and
Eq.~(\ref{eq:wave4}) for $\phi_{g}$ by $\phi_{n}$ and subtract the
obtained two
equations.
Then we have 
\begin{equation}
 \left\{h\left(\phi_{g}\phi_{n,x}-\phi_{n}\phi_{g,x}\right)\right\}_{,x}+\frac{(\gamma_{n}-\gamma_{g})\sqrt{\alpha}x}{V'(1-x^{2})}\left\{\frac{2\phi_{g,x}}{\phi_{g}}+\frac{\sqrt{\alpha}x(F+W)}{V'(1-x^{2})}\right\}h\phi_{g}\phi_{n}=0~,\label{eq:varphi}
\end{equation}
where $\gamma_{g}$ is the imaginary part of $\omega_{g}$, which is given
by Eq.~(\ref{eq:omegag}).
By virtue of the boundary condition at the regular center, the
integration of Eq.~(\ref{eq:varphi}) leads to 
\begin{equation}
 h\left(\phi_{g}\phi_{n,x}-\phi_{n}\phi_{g,x}\right)=-(\gamma_{n}-\gamma_{g})\int_{0}^{x}\frac{\sqrt{\alpha}x}{V'(1-x^{2})}\left\{\frac{2\phi_{g,x}}{\phi_{g}}+\frac{\sqrt{\alpha}x(F+W)}{V'(1-x^{2})}\right\}h\phi_{g}\phi_{n} dx~.\label{eq:varphix}
\end{equation}

Now let us turn our attention to the asymptotic behaviors of the both
sides of Eq.~(\ref{eq:varphix}) near $x=1$.
Because of Eq.~(\ref{eq:hs}), the left hand side of
Eq.~(\ref{eq:varphix}) near $x=1$ is given by  
\begin{equation}
 h\left(\phi_{g}\phi_{n,x}-\phi_{n}\phi_{g,x}\right) = C_{5}(\omega)(1-x)^{-a} + O\left((1-x)^{-a+1}\right).\label{eq:ex}
\end{equation}
If $\gamma_{n}$ is in the range $1+(V'_{s}/\sqrt{\alpha})<\gamma_{n}<1$
(i.e., $0<a<1$), the subleading term proportional to $(1-x)^{-a+1}$ must
vanish in the limit $x\rightarrow 1$.
This is a key point to discuss the absence of non-oscillatory normal
modes with $\gamma_{n}$ giving the range $0<a<1$ by comparing
Eq.~(\ref{eq:ex}) with the right hand side of Eq.~(\ref{eq:varphix}). 

Let us introduce the function $H$ defined as
\begin{equation}
 H(x) \equiv \frac{\sqrt{\alpha}x(1-x)^{a}}{V'(1+x)}\left\{\frac{2\phi_{g,x}}{\phi_{g}}+\frac{\sqrt{\alpha}x(F+W)}{V'(1-x^{2})}\right\}h\phi_{g}~,
\end{equation}
which is the function defined by the background solution.
It becomes zero at $x=0$ and a
finite value (denoted by $H_{s}$) at $x=1$.
Using the function $H$, we can rewrite the integral in the right hand
side of Eq.~(\ref{eq:varphix}) into the form
\begin{eqnarray}
 &&\int_{0}^{x}\frac{\sqrt{\alpha}x}{V'(1-x^{2})}\left\{\frac{2\phi_{g,x}}{\phi_{g}}+\frac{\sqrt{\alpha}x(F+W)}{V'(1-x^{2})}\right\}h\phi_{g}\phi_{n}
  dx \nonumber\\
&&=\frac{H_{s}}{a}(1-x)^{-a}\phi_{n}-\frac{H_{s}}{a}\int_{0}^{x}(1-x)^{-a-1}\left\{(1-x)\phi_{n,x}-a\left(\frac{H}{H_{s}}-1\right)\phi_{n}\right\}dx~.\label{eq:partial}
\end{eqnarray}
For $0<a<1$, the integral in the second term of the right hand side of
Eq.~(\ref{eq:partial}) must vanish in the limit $x\rightarrow1$ for
consistency between Eq.~(\ref{eq:ex}) and Eq.~(\ref{eq:partial}).
Thus we obtain the condition for non-oscillatory normal modes
with $\gamma_{n}$ giving $0<a<1$ as follows, 
\begin{equation}
 \int_{0}^{1}(1-x)^{-a-1}\left\{(1-x)\phi_{n,x}-a\left(\frac{H}{H_{s}}-1\right)\phi_{n}\right\}dx=0~.\label{eq:condition}
\end{equation}

It is clear that the condition (\ref{eq:condition}) cannot be satisfied
if $H/H_{s}<1$ and $\phi_{n,x}/\phi_{n}>0$ for any $x$ in the range
$0\leq x \leq 1$.
Fortunately, we can show that if Eq.~(\ref{eq:ineq}) is satisfied, the positivity of $\phi_{n,x}/\phi_{n}$ is
assured.
To see this, we firstly note that Eq.~(\ref{eq:wave4}) requires that if
$\phi_{n}$ becomes locally maximum at $m$ points $x=x_{i}$ (satisfying
$0<x_{1}<\cdots<x_{m}<1$), the ratio $\phi_{n,xx}/\phi_{n}$ must be
positive there.
However, recalling that $\phi_{n}\sim x$ near $x=0$, we can claim that
the ratio $\phi_{n,xx}/\phi_{n}$ must be negative at $x=x_{1}$.
This contradiction is caused by the assumption of the existence of the
points at which $\phi_{n,x}=0$.
Hence we can conclude that if a background self-similar system satisfies
both the conditions $H/H_{s}<1$ and (\ref{eq:ineq}) for any $x$ in the
range $0\leq x \leq 1$, there exists no normal mode with $\gamma_{n}$
giving $0<a<1$.

It is interesting to check whether the flat Friedmann solution can
satisfy such conditions.
Because the ratio $H_{,x}/H$ is given by
\begin{equation}
 \frac{H_{,x}}{H}=\frac{6(1-\alpha)x^{2}+10(1+3\alpha)x+3(1+3\alpha)^{2}-3(1+\alpha)x(1+3\alpha+2x)\gamma_{n}}{(1+3\alpha)x(1+x)(1+3\alpha+2x)}~,
\end{equation}
it is easy to understand that if $\gamma_{n}<1$, the function $H$ has no
extremum in the range $0\leq x \leq 1$, and hence we obtain $H/H_{s}<1$ there.
Further we can find that for $\gamma_{n}>0$ the second derivative of
$U-\gamma_{n}(F+W)$ with respect to $x$ become always positive in the
subsonic region.
Hence, the positivity of $U-\gamma_{n}(F+W)$ in the subsonic region is
assured if its derivative with respect to $x$ is negative at $x=1$,
which leads to  
\begin{equation}
 \gamma_{n} \geq q \equiv \frac{(1+3\alpha)(1-9\alpha)}{9\alpha^{2}+18\alpha+1}~.\label{eq:b}
\end{equation}
Note that the value of $1+(V'_{s}/\sqrt{\alpha})$ for the flat Friedmann
solution is equal to $p$, which is given in Eq.~(\ref{eq:p}), and that
the equality $p=q$ holds at $\alpha=\alpha_{c}\approx 0.01879$.
Thus the absence of non-oscillatory normal modes with $\gamma_{n}$ lying
in
the range $\gamma_{n}>p$ is assured only for 
$\alpha_{c} \leq \alpha\leq 1$, 
while for $0<\alpha\leq\alpha_{c}$, the validity of the proof of the
absence holds for non-oscillatory normal modes with the range $\gamma_{n}>q$.
Further developments of methods to analyze non-oscillatory normal modes with the smaller range of $\gamma_{n}$
will be necessary for claiming definitely the absence of any
non-oscillatory unstable normal modes.
However they require more complicated and sophisticated techniques
to analyze Eq.~(\ref{eq:wave4}), which also remains to be investigated
in future works.

\section{Summary and discussion\label{sec:sum}}

Now let us summarize the results obtained in this paper.
We have considered a spherically symmetric system describing gravitational collapse of a perfect fluid with the equation of state
$P=\alpha\rho$.
To treat analytically time evolution of the perturbations in the
self-similar background (which may describe bouncing and recollapsing
motion of a fluid), we have constructed the single wave equation governing the
perturbations.
As an application of the wave equation to the stability analysis, the
eigenvalue problem for the perturbations having the time dependence
$\exp{\{i\omega\log(-t)\}}$ has been studied, and we have arrived at the
following main conclusion:
Non-oscillatory ($\beta_{n}=0$) unstable normal modes with the growth
rate $\gamma_{n}>1$ do not exist if Eq.~(\ref{eq:ineq}) holds for the
self-similar backgrounds in the subsonic region, and some
additional conditions such that $H/H_{s}<1$ become necessary for the
absence of non-oscillatory unstable normal modes with $\gamma_{n}<1$.

In this paper we have found the absence of non-oscillatory unstable normal modes,
except for ones with the small growth rate such as $\gamma_{n}<q$ (for
$0<\alpha\leq\alpha_{c}$) or $\gamma_{n}<p$ (for $\alpha_{c}\leq\alpha\leq 1$), in
the flat Friedmann background for $0<\alpha\leq 1$.
This supports the numerical result given by \cite{HaradaT:PRD63:2001}
for $0<\alpha\lesssim 0.036$.
Because the attractor behavior of the flat Friedmann solution was also pointed out in
\cite{HaradaT:PRD63:2001}, the absence of oscillatory unstable normal
modes with nonzero $\beta_{n}$ may be also shown in our analytical
scheme.
In contrast to the case of the flat Friedmann background, it has been numerically found that a non-oscillatory
unstable normal mode can be excited in the critical gravitational
collapse for $0<\alpha\lesssim 0.89$ \cite{KoikeT:PRD59:1999}.
The consistency of the numerical result with our analytical scheme may
be checked by showing that the condition given by Eq.~(\ref{eq:ineq})
breaks down for the critical solution.

Finally we would like to remark that the absence of non-oscillatory
unstable normal modes with $\gamma_{n}>1$ is analytically shown for the Newtonian
Larson-Penston solution \cite{HanawaT:AJ484:1997}.
It is interesting to note that some mathematical difficulties also appear
in the Newtonian analysis of the slowly growing modes with
$\gamma_{n}<1$.
The result obtained by the normal mode analysis for the Newtonian Larson-Penston background strongly
suggests that the condition (\ref{eq:ineq}) (for $\gamma_{n}>1$) should
be satisfied for its general relativistic version.

\appendix

\section{Asymptotic behavior of self-similar solutions near the sonic
 point\label{sec:sonic}}

From Eqs.~(\ref{eq:Veq}), (\ref{eq:Xeq})
and (\ref{eq:A''}) we obtain the expansion of the functions $V$, $A'$ and $X$ near the
sonic point $\xi=\xi_{s}$ as
\begin{eqnarray}
 V(\xi) &=&
  -\sqrt{\alpha}+\sum_{i=1}^{\infty}V_{i}(\xi-\xi_{s})^{i}+(\xi-\xi_{s})^{\chi}\sum_{j=0}^{\infty}V_{\chi+j}(\xi-\xi_{s})^{j}~,\label{eq:Vex}\\
 A'(\xi) &=&
  \sum_{i=0}^{\infty}A_{i}'(\xi-\xi_{s})^{i}+(\xi-\xi_{s})^{\chi}\sum_{j=0}^{\infty}A_{\chi+j}'(\xi-\xi_{s})^{j}~,\label{eq:A'ex}\\
X(\xi) &=&
  \sum_{i=0}^{\infty}X_{i}(\xi-\xi_{s})^{i}+(\xi-\xi_{s})^{\chi}\sum_{j=0}^{\infty}X_{\chi+j}(\xi-\xi_{s})^{j}~,\label{eq:Xex}
\end{eqnarray}
where the variable $\xi$ is defined as $\xi\equiv\log(-z)$, and the
power index $\chi$ is given by
\begin{eqnarray}
 V_{1}\chi& =&
  -\frac{(1+\alpha)(1+3\alpha)A'_{0}+1+5\alpha}{2\sqrt{\alpha}}\nonumber\\
&&-\frac{\left\{(1+\alpha)^{2}(1+3\alpha)A^{\prime2}_{0}+2(1+\alpha)(1+5\alpha)A'_{0}+\alpha^{2}+6\alpha+1\right\}X_{\chi}}{4\sqrt{\alpha}(1+\alpha)X_{0}A'_{\chi}}\nonumber\\
&&-\frac{V_{\chi}A_{1}'}{A_{\chi}'}-\frac{2V_{\chi}A'_{0}(1-\alpha)\left\{3(1+\alpha)A'_{0}+2\right\}}{4(1+\alpha)A'_{\chi}}~.\label{eq:chieq}
\end{eqnarray}
We also obtain
\begin{eqnarray}
 &&
  \frac{V_{\chi}}{A'_{\chi}}=-\frac{\sqrt{\alpha}(1-\alpha)}{1+(1+\alpha)A'_{0}}~,\label{eq:Vchieq}\\
 &&\frac{X_{\chi}}{A'_{\chi}}=-\frac{2X_{0}}{1+(1+\alpha)A'_{0}}~,\label{eq:Xchieq}\\
 &&X_{0}=\frac{(1+\alpha)^{2}}{(1+\alpha)^{2}(3\alpha+1)A_{0}^{\prime2}+2(1+\alpha)(1+5\alpha)A_{0}'+\alpha^{2}+6\alpha+1}~.\label{eq:X0}
\end{eqnarray}
Moreover we can obtain the quadratic equation for $A'_{1}$ and find
\begin{equation}
 A'_{1}=A_{1\pm}' \equiv \frac{\left(1+2A_{0}'\right)\left\{(1+\alpha)(5\alpha-1)A_{0}'+3\alpha-1\right\}\pm\left\{1+(1+\alpha)A_{0}'\right\}\sqrt{y}}{2(1-\alpha^{2})}~,\label{eq:A'1}
\end{equation}
where $y$ is defined by
\begin{equation}
 y=4(1+\alpha)^{2}(1+3\alpha)A_{0}^{\prime2}+4(1+\alpha)(1+3\alpha^{2})A_{0}'+(1-3\alpha)^{2}~.
\end{equation}
Then $V_{1}$ is written by
\begin{equation}
 V_{1}=V_{1\pm}\equiv -\frac{\sqrt{\alpha}}{2}\left(1 \pm \frac{\sqrt{y}}{1+\alpha}\right)~.\label{eq:V1}
\end{equation}
Substituting Eqs.~(\ref{eq:Vchieq}), (\ref{eq:Xchieq}), (\ref{eq:X0})
and (\ref{eq:A'1}) to Eq.~(\ref{eq:chieq}), we arrive at the simple form
\begin{equation}
 \chi = -1-\frac{\sqrt{\alpha}}{V_{1}}~.
\end{equation}
The solutions $V$, $A'$ and $X$ are $C^{1}$ or analytic at the sonic point when
either $\chi\geq 1$ or $A'_{\chi}=0$ (i.e., the second terms of
Eqs.~(\ref{eq:Vex}), (\ref{eq:A'ex}) and (\ref{eq:Xex}) disappear).
Then, it is easy to see that $V_{1}=V'_{s}$, $A_{0}=A'_{s}$, $A_{1}=A''_{s}$,
$X_{0}=X_{s}$ and $X_{1}=X'_{s}$.

\section{Functions in the perturbation equations\label{sec:functions}}
We can write the coefficient functions $P_{i}$ and $Q_{i}$ in
Eqs.~(\ref{eq:P}) and (\ref{eq:Q}) as
\begin{eqnarray}
 P_{1}&=&\frac{2X}{1+\alpha}\left\{1+(1+\alpha)A'\right\}\left\{\alpha
									     V^{2}A^{\prime2}-\left(1+A'\right)^{2}\right\}~,\\
P_{2}&=&-\frac{1+A'}{1+\alpha}\left[1+\alpha+X(1-\alpha)\left\{V^{2}A^{\prime2}+\left(1+A'\right)^{2}\right\}\right]~,\\
P_{3}&=&-2X(1+A')A'V^{2}~,\\
P_{4}&=&-\frac{2X}{1+\alpha}\left\{-V^{2}A'\left(1+\alpha+A'\right)+\alpha\left(1+A'\right)^{2}\right\}~,\\
P_{5}&=&\frac{1+A'}{1+\alpha}\left[1+\alpha+X\left\{3(1-\alpha)V^{2}A^{\prime2}-(1+7\alpha)\left(1+A'\right)^{2}\right\}\right]~,\\
Q_{2}&=&\frac{\alpha A'}{1+A'}P_{2}~,\\
Q_{3}&=&-\frac{2\alpha
 X}{1+\alpha}\left\{V^{2}A^{\prime2}+\left(1+A'\right)^{2}\right\}~,\\
Q_{4}&=&-\frac{2\alpha
 X}{1+\alpha}\left\{-V^{2}A^{\prime2}+\left(1+A'\right)\left(\alpha
									      A'-1\right)\right\}~,\\
Q_{5}&=&\frac{\alpha A'}{1+A'}P_{5}~.
\end{eqnarray}
The coefficient function $T_{i}$ in Eq.~(\ref{eq:T}) are given
by
\begin{eqnarray}
 T_{1}& =& -\left(P_{3}Q_{4}-P_{4}Q_{3}\right)f+P_{1}\left(P_{3}+Q_{3}\right)=-2XA'\left(1+A'\right)
  P_{1}V\left(V\pm\sqrt{\alpha}\right)~,\\
 T_{2}
  &=&P_{1}\left(Q_{4}+P_{4}\right)=2XA'\left(1+A'\right)P_{1}\left(V^{2}-\alpha\right)~,\\
T_{3}&=&-\left(P_{5}Q_{4}-P_{4}Q_{5}\right)f+P_{1}\left(P_{5}+Q_{5}\right)=
  \frac{(1+\alpha)P_{1}P_{5}A'\left(V\pm\sqrt{\alpha}\right)}{A'V\pm\sqrt{\alpha}\left(1+A'\right)}~,
\end{eqnarray}
where the upper and lower sign correspond to the choice of $f=f_{+}$ and
$f=f_{-}$, respectively.
In addition the function $Y$ is given by
\begin{equation}
 Y=\left(P_{5}Q_{4}-P_{4}Q_{5}\right)f^{2}+\left(P_{2}Q_{4}-P_{4}Q_{2}-P_{1}P_{5}-P_{1}Q_{5}\right)f-P_{1}\left(Q_{4}+P_{4}\right)f'-P_{1}\left(P_{2}+Q_{2}\right)~.\label{eq:Y}
\end{equation}
Using Eqs.~(\ref{eq:Veq}) and (\ref{eq:A''}), we write $A''$ and $V'$ in
$f'$ of Eq.~(\ref{eq:Y}) by $V$ and $A'$ and $X$ to obtain a simpler form
of $Y$.
Then we have
\begin{equation}
 Y=\frac{P_{1}A'\left(1+A'\right)V\left(Y_{1}X+Y_{2}\right)}{\sqrt{\alpha}\left\{A'V\pm\sqrt{\alpha}\left(1+A'\right)\right\}^{2}}~,
\end{equation}
where the functions $Y_{1}$ and $Y_{2}$ are written as
\begin{eqnarray}
 Y_{1}&=&\pm\left\{(1+\alpha)^{2}(3\alpha+1)A^{\prime2}+(-3\alpha^{3}+9\alpha^{2}+15\alpha+3)A'+2(-\alpha^{2}+4\alpha+1)\right\}A^{\prime2}V^{2}\nonumber\\
&&+8\alpha\sqrt{\alpha}\left(1+A'\right)\left\{1+(1+\alpha)A'\right\}A'V\nonumber\\
&&\mp\left(1+A'\right)^{2}\left\{(1+\alpha)^{2}(3\alpha+1)A^{\prime2}+(1+\alpha)(\alpha^{2}+16\alpha+3)A'+2(\alpha^{2}+6\alpha+1)\right\}~,\nonumber\\ \\
Y_{2}&=&\pm(1+\alpha)^{2}\left\{(1-\alpha)A^{\prime2}+(3+\alpha)A'+2\right\}~.
\end{eqnarray}

The functions $W$, $F$ and $U$ in Eq.~(\ref{eq:wave}) are
given by
\begin{eqnarray}
 W&=&-\frac{V^{2}-\alpha}{\alpha}\left\{-\frac{2\alpha
					 V'}{V(V^{2}-\alpha)}+\frac{K_{1}}{J}+\frac{V^{2}K_{2}}{J(V^{2}-\alpha)}\right\}~,\\
F&=&-\frac{V}{\sqrt{\alpha}}\left\{-\frac{\left(V^{2}+\alpha\right)V'}{V^{3}}+\frac{K_{2}}{J}\right\}~,\\
U&=&-\frac{\left(V^{2}-\alpha\right)K_{3}}{\alpha J}~,
\end{eqnarray}
where the functions $K_{1}$, $K_{2}$, $K_{3}$ and $J$ are written as
\begin{eqnarray}
 K_{1}&=&\left(P_{1}+P_{4}f\right)\left(YT_{1}'-Y'T_{1}\right)+\left(P_{2}+P_{4}f'+P_{5}f\right)YT_{1}+YP_{3}fT_{3}+P_{3}Y^{2}~,\\
  K_{2}&=&\left(P_{1}+P_{4}f\right)\left(YT_{2}'-Y'T_{2}+YT_{3}\right)+\left(P_{2}+P_{4}f'+P_{5}f\right)YT_{2}+Y^{2}P_{4}~,\\
K_{3}&=&\left(P_{1}+P_{4}f\right)\left(YT_{3}'-Y'T_{3}\right)+\left(P_{2}+P_{4}f'+P_{5}f\right)YT_{3}+Y^{2}P_{5}~,\\
 J &=& YP_{3}fT_{1}~.
\end{eqnarray}
Using Eqs.~(\ref{eq:Veq}), (\ref{eq:Xeq}) and (\ref{eq:A''}), we can
also write the functions $W$, $F$ and $U$ by $V$, $A'$ and $X$ only.
Then we have
\begin{eqnarray}
 W &=&
  -\frac{W_{1}X^{2}+W_{2}X+W_{3}}{\sqrt{\alpha}(1+\alpha)VA'\left(1+A'\right)\left(Y_{1}X+Y_{2}\right)X}~,\\
F&=&\mp
 W\pm\frac{F_{1}X+F_{2}}{2\sqrt{\alpha}(1+\alpha)A'\left(1+A'\right)V\left\{1+(1+\alpha)A'\right\}\left\{A'V\pm\sqrt{\alpha}\left(1+A'\right)\right\}X}~,\nonumber\\ \\
U&=&\frac{T_{3}}{T_{4}}W\pm\frac{U_{1}X^{2}+U_{2}X+U_{3}}{2\alpha(1+\alpha)A'\left(1+A'\right)V^{2}\left\{1+(1+\alpha)A'\right\}\left\{A'V\pm\sqrt{\alpha}\left(1+A'\right)\right\}X^{2}}~,\nonumber\\
\end{eqnarray}
where the functions $W_{1}$, $W_{2}$, $W_{3}$, $F_{1}$, $F_{2}$,
$U_{1}$, $U_{2}$, $U_{3}$ and $T_{4}$ are given by 
\begin{eqnarray}
 W_{1}&=&(1+3\alpha)\left\{-2(1+\alpha)^{3}A^{\prime3}+(\alpha-3)(1+\alpha)^{2}(3\alpha+2)A^{\prime2}\right.\nonumber\\
&&\hspace{3cm}\left.+2(\alpha-3)(1+\alpha)(1+3\alpha)A'+2(\alpha^{2}-4\alpha-1)\right\}A^{\prime4}V^{4}\nonumber\\
&&\pm2\sqrt{\alpha}(1+\alpha)\left(1+A'\right)\left\{-(1+\alpha)^{2}(1+3\alpha)A^{\prime2}\right.\nonumber\\
&&\hspace{3cm}\left.+3(1+\alpha)(\alpha^{2}-4\alpha-1)A'+2(\alpha^{2}-4\alpha-1)\right\}A^{\prime3}V^{3}\nonumber\\
&&+(1+\alpha)\left(1+A'\right)^{2}\left\{2(1+\alpha)^{3}(1+3\alpha)A^{\prime3}-(1+\alpha)^{2}(3\alpha-5)(1+3\alpha)A^{\prime2}\right.\nonumber\\
&&\hspace{3cm}\left.+4(-11\alpha^{3}-9\alpha^{2}+3\alpha+1)A'-\alpha^{3}-33\alpha^{2}+\alpha+1\right\}A^{\prime2}V^{2}\nonumber\\
&&\pm2\sqrt{\alpha}(1+\alpha)\left(1+A'\right)^{3}\left\{(1+\alpha)^{2}\left(1+3\alpha\right)A^{\prime2}\right.\nonumber\\&&\hspace{3cm}\left.+(1+\alpha)(\alpha^{2}+16\alpha+3)A'+2(\alpha^{2}+6\alpha+1)\right\}A'V\nonumber\\
&&-\left(1+A'\right)^{4}\left\{2\alpha(1+\alpha)^{3}(1+3\alpha)A^{\prime3}-(1+\alpha)^{2}(1+3\alpha)(1+5\alpha)A^{\prime2}\right.\nonumber\\
&&\hspace{3cm}\left.-2(1+\alpha)(1+5\alpha)(\alpha^{2}+6\alpha+1)A'-(\alpha^{2}+6\alpha+1)^{2}\right\}~,
\end{eqnarray}
\begin{eqnarray}
 W_{2}&=&-(1+\alpha)^{2}\left[\left\{2(1+\alpha)^{2}A^{\prime3}+(6\alpha^{3}+7\alpha^{2}+10\alpha+5)A^{\prime2}\right.\right.\nonumber\\
&&\hspace{5cm}\left.+4(1+\alpha)(1+2\alpha)A'-\alpha^{2}+4\alpha+1\right\}A^{\prime2}V^{2}\nonumber\\
&&\pm2\sqrt{\alpha}(1+\alpha)\left(1+A'\right)\left\{(1-\alpha)A^{\prime2}+(3+\alpha)A'+2\right\}A'V\nonumber\\
&&-\left(1+A'\right)^{2}\left\{2\alpha(1+\alpha)^{2}A^{\prime3}-(\alpha^{3}+6\alpha^{2}+11\alpha+2)A^{\prime2}\right.\nonumber\\
&&\hspace{5cm}\left.\left.-4(3\alpha^{2}+6\alpha+1)A'-2(\alpha^{2}+6\alpha+1)\right\}\right]~,
\end{eqnarray}
\begin{eqnarray}
W_{3}&=&(1+\alpha)^{4}\left\{(1+\alpha)A^{\prime2}+2A'+1\right\}~,\\
F_{1}&=&\left\{(1+\alpha)^{2}(9\alpha-5)A^{\prime2}+3(1+\alpha)(\alpha^{2}+6\alpha-3)A'+4(\alpha^{2}+2\alpha-1)\right\}A^{\prime3}V^{3}\nonumber\\
&&+\sqrt{\alpha}(1+\alpha)\left(1+A'\right)\left\{(1+\alpha)(3\alpha-7)A^{\prime2}+(3\alpha^{2}+2\alpha-9)A'+2(\alpha-1)\right\}A^{\prime2}V^{2}\nonumber\\
&&+(1+\alpha)\left(1+A'\right)^{2}\left\{(1+\alpha)(11\alpha+1)A^{\prime2}+(5\alpha^{2}+14\alpha+1)A'+4\alpha\right\}A'V\nonumber\\
&&+\sqrt{\alpha}\left(1+A'\right)^{3}\left\{(1+\alpha)^{2}(17\alpha+3)A^{\prime2}+(1+\alpha)(\alpha^{2}+30\alpha+5)A'+2(\alpha^{2}+6\alpha+1)\right\}~,\nonumber\\ \\
F_{2}&=&-(1+\alpha)^{2}\left[\left\{(1+3\alpha)A'+1+\alpha\right\}A^{\prime2}V+\sqrt{\alpha}\left(1+A'\right)^{2}\left\{(3+\alpha)A'+2\right\}\right]~,
\end{eqnarray}
\begin{eqnarray}
U_{1}&=&\pm3(1-\alpha)\left\{(1+\alpha)^{2}(1+3\alpha)A^{\prime2}+2(1+\alpha)(1+5\alpha)A'+\alpha^{2}+6\alpha+1\right\}A^{\prime4}V^{5}\nonumber\\
&&+6\sqrt{\alpha}(1-\alpha^{2})\left(1+A'\right)\left\{1+(1+\alpha)A'\right\}A^{\prime3}V^{4}\nonumber\\
&&\pm3(\alpha-1)\left(1+A'\right)\left\{(1+\alpha)^{2}(1+3\alpha)A^{\prime3}+(1+\alpha)(\alpha^{2}+12\alpha+3)A^{\prime2}\right.\nonumber\\
&&\hspace{7cm}\left.+(9\alpha^{2}+16\alpha+3)A'+\alpha^{2}+6\alpha+1\right\}A^{\prime2}V^{3}\nonumber\\
&&+\sqrt{\alpha}(1+7\alpha)\left(1+A'\right)^{2}\left\{(1+\alpha)^{2}(1+3\alpha)A^{\prime3}+8\alpha(1+\alpha)A^{\prime2}\right.\nonumber\\
&&\hspace{8cm}\left.-(\alpha^{2}+3)A'-2(1+\alpha)\right\}A'V^{2}\nonumber\\
&&\mp2\alpha(1+\alpha)(1+7\alpha)\left(1+A'\right)^{3}\left\{1+(1+\alpha)A'\right\}A'V\nonumber\\
&&-\sqrt{\alpha}(1+7\alpha)\left(1+A'\right)^{4}\left\{(1+\alpha)^{2}(1+3\alpha)A^{\prime2}\right.\nonumber\\
&&\hspace{7cm}\left.+2(1+\alpha)(1+5\alpha)A'+\alpha^{2}+6\alpha+1\right\}~,
\end{eqnarray}
\begin{eqnarray}
 U_{2}&=&-(1+\alpha)\left[\mp\left\{3(1+\alpha)(2\alpha^{2}-\alpha+1)A^{\prime2}+2(1+\alpha)(3+\alpha)A'-3\alpha^{2}+4\alpha+3\right\}A^{\prime2}V^{3}\right.\nonumber\\
&&+\sqrt{\alpha}(1+\alpha)\left(1+A'\right)\left\{(1+\alpha)A^{\prime2}-(1+\alpha)A'-2\right\}A'V^{2}\nonumber\\
&&\mp\alpha(1+\alpha)\left(1+A'\right)\left\{(1+\alpha)A^{\prime2}+3(1+\alpha)A'+2\right\}A'V\nonumber\\
&&+\sqrt{\alpha}\left(1+A'\right)^{2}\left\{(1+\alpha)(3\alpha^{2}-11\alpha-2)A^{\prime2}\right.\nonumber\\
&&\hspace{6cm}\left.\left.-4(1+\alpha)(1+6\alpha)A'-2(4\alpha^{2}+7\alpha+1)\right\}\right]~,\\
U_{3}&=&\mp\sqrt{\alpha}(1+\alpha)^{3}\left\{\sqrt{\alpha}A^{\prime2}V\pm\left(1+A'\right)^{2}\right\}~,\\
T_{4}&=& \mp2\sqrt{\alpha}VXA'\left(1+A'\right)P_{1}~.
\end{eqnarray}

\bibliography{refgrqc}
\end{document}